\RequirePackage{rotating}
\documentclass[fleqn,usenatbib,useAMS]{mnras} 

\usepackage[T1]{fontenc}
\usepackage{ae,aecompl}

\usepackage{graphicx}	
\usepackage{amsmath}	
\usepackage{amssymb}	
\usepackage{cite}
\usepackage{natbib}
\usepackage{subfigure}
\usepackage{pdflscape}
\usepackage{hyperref}

\title[Abundance ratios in dwarf elliptical galaxies]{Abundance ratios in dwarf elliptical galaxies}

\author[\c{S}. \c{S}en et al.]{
\c{S}. \c{S}en$^{1,2}$\thanks{E-mail: s.sen@rug.nl},
R. F. Peletier$^{1}$,
A. Boselli$^{3}$,
M. den Brok$^{4}$,
J. Falc{\'o}n-Barroso$^{5,6}$, 
\newauthor G. Hensler$^{7}$, J. Janz$^{8,9,16}$, E. Laurikainen$^{9}$, T. Lisker$^{10}$, J. J. Mentz$^{1,11}$, S. Paudel$^{12}$,  
\newauthor H. Salo$^{9}$, A. Sybilska$^{13}$, E. Toloba$^{14}$, G. van de Ven$^{15}$, A. Vazdekis$^{5,6}$ and C. Yesilyaprak$^{2}$
\\
\\
$^{1}$Kapteyn Astronomical Institute, University of Groningen, P. O. Box 800, 9700 AV Groningen, Netherlands\\
$^{2}$Dept. of Astronomy and Astrophysics, Faculty of Science, Atat\"urk University, 25030, Erzurum, Turkey\\
$^{3}$Aix Marseille Universit{\'e}, CNRS, LAM (Laboratoire d'Astrophysique de Marseille), UMR 7326, 13388 Marseille, France\\
$^{4}$Institute for Astronomy, ETH Zurich, Wolfgang-Pauli-Strasse 27, 8093 Zurich, Switzerland\\
$^{5}$Instituto de Astrof{\'i}sica de Canarias,Calle V{\'i}a L{\'a}ctea s/n, E-38200 La Laguna, Tenerife, Spain\\
$^{6}$Departamento de Astrof\'isica, Universidad de La Laguna, E-38205 La Laguna, Tenerife, Spain\\
$^{7}$Department of Astrophysics, University of Vienna, T\"urkenschanzstrasse 17, 1180 Vienna, Austria\\
$^{8}$Centre for Astrophysics and Supercomputing, Swinburne University, Hawthorn, VIC 3122, Australia\\
$^{9}$Astronomy Research Unit, University of Oulu, FI-90014, Finland\\
$^{10}$Astronomisches Rechen-Institut, Zentrum f\"ur Astronomie der Universit\"at Heidelberg, M\"onchhofstra{\ss}e 12-14,69120 Heidelberg, Germany\\
$^{11}$Center for Space Research, North-west University, Potchefstroom 2520, South Africa\\
$^{12}$Department of Astronomy \& Center for Galaxy Evolution Research, Yonsei University, Seoul 03722, Republic of Korea\\
$^{13}$European Southern Observatory, Karl-Schwarzschild-Strasse 2, 85748 Garching bei M\"unchen, Germany\\
$^{14}$University of the Pacific, Department of Physics, 3601 Pacific Avenue, Stockton, CA 95211, USA \\
$^{15}$Max-Planck-Institut f\"ur Astronomie, K\"onigsstuhl 17, 69117 Heidelberg, Germany \\
$^{16}$Finnish Centre of Astronomy with ESO (FINCA), University of Turku, V\"ais\"al\"antie 20, 21500 Piikki\"o, Finland
}

\date{Accepted XXX. Received YYY; in original form ZZZ}

\pubyear{2017}

\begin{document}
\label{firstpage}
\pagerange{\pageref{firstpage}--\pageref{lastpage}}
\maketitle

\begin{abstract}
We determine abundance ratios of 37 dwarf
ellipticals (dEs) in the nearby Virgo cluster. This sample is representative of the early-type population of galaxies in the absolute magnitude range -19.0 < M$_r$ < -16.0. We analyze their absorption line-strength indices by 
means of index-index diagrams and scaling relations and use the stellar population models to interpret them. We present ages, metallicities and abundance ratios obtained from these dEs within an aperture size of ${R}_{e}$/8. We
calculate [Na/Fe] from NaD, [Ca/Fe] from Ca4227 and [Mg/Fe] from Mgb. We find that [Na/Fe] is under-abundant with respect to solar
while [Mg/Fe] is around solar. This is exactly opposite to what is found for
giant ellipticals, but follows the trend with metallicity found previously for the
Fornax dwarf NGC 1396. We discuss possible formation scenarios that can result in such elemental abundance patterns and we speculate that dEs have disk-like SFH favouring them to originate from late-type dwarfs or small spirals. 
Na-yields appear to be very metal-dependent, in agreement with studies of giant ellipticals, probably due to the large dependence on the neutron-excess in stars. We conclude that dEs have undergone a considerable amount of chemical evolution, they are therefore not uniformly old, but have extended SFH, similar to many of the Local Group galaxies. 

\end{abstract}




\section{Introduction}
Early-type dwarf galaxies (dEs) play a key role in understanding galaxy cluster evolution. dEs\footnote[16]{The term dE has traditionally been used to refer to dwarf elliptical galaxies, whereas we loosely use the term here to include dwarf ellipticals and dwarf lenticulars (dS0).}, the low
luminosity (${M}_\text{B}$ > -18) and low surface brightness ($\mu_\text{B}$ > 22
{\,mag\,arcsec$^{-2}$}) population of the Early Type Galaxy (ETGs) class are found in high-density
environments and are very rare in isolation \citep{2010AA...517A..73G,2012ApJ...757...85G,blantonetal.2005ApJ...629..143B}. dEs are found
abundantly in groups and clusters of galaxies where they dominate in numbers \citep{binggelietal.1988ARA&A..26..509B}. 

The Lambda cold dark matter (CDM) hierarchical merging scenario predicts that CDM haloes are formed because
of gravitational instabilities and evolve hierarchically via mergers (\citealt*{1978MNRAS.183..341W};
\citealt{1988ApJ...327..507F}; \citealt*{1991ApJ...379...52W,1993ASPC...51...192L};
\citealt{2000MNRAS.319..168C}). These models predict that dwarf-size dark matter haloes form first and then
merge forming more massive haloes.

The class of dEs contains objects covering a wide range of internal properties, with sometimes rather
complicated structures. Taking advantage of deep photometric studies, we know that several of them contain
substructures such as disks, sprial arms and irregular features
(e.g. \citealt{jerjenetal.2000A&A...358..845J,barazzaetal.2002A&A...391..823B,gehaetal.2003AJ....126.1794G}; \citealt*{grahamandguzman.2003AJ....125.2936G}; \citealp{derijckeetal.2003A&A...400..119D,liskeretal.2006AJ....132..497L,ferrareseetal.2006ApJS..164..334F,janzetal.2012ApJ...745L..24J,janzetal.2014ApJ...786..105J}). Apart from this, dEs also show a complicated
variety of internal kinematics and dynamics. dEs with similar photometric properties can have different
stellar populations (\citealt{michielsenetal.2008MNRAS.385.1374M,paudeletal.2010MNRAS.405..800P,kolevaetal.2009MNRAS.396.2133K,kolevaetal.2011MNRAS.417.1643K,rysetal.2015MNRAS.452.1888R}) and different rotation speeds
(\citealt{pedrazetal.2002MNRAS.332L..59P}; \citealt*{simienandprugniel.2002A&A...384..371S}; \citealt{gehaetal.2002AJ....124.3073G,gehaetal.2003AJ....126.1794G,vanzeeetal.2004AJ....128.2797V,chilingarian.2009MNRAS.394.1229C,tolobaetal.2009ApJ...707L..17T,tolobaetal.2011A&A...526A.114T,tolobaetal.2014ApJ...783..120T,tolobaetal.2015ApJ...799..172T,kolevaetal.2009MNRAS.396.2133K,kolevaetal.2011MNRAS.417.1643K,rysetal.2013MNRAS.428.2980R,rysetal.2014MNRAS.439..284R}). \citet{kormendy.1985ApJ...295...73K} suggested that they developed their spheroidal non-star forming
appearance, that is probably highly flattened
(\citealt{liskeretal.2006AJ....132..497L,liskeretal.2007ApJ...660.1186L}), during a transformation from a
late-type galaxy that fell into a cluster; it is thought that this transformation is induced by the
environment because the morphology-density relation largely depends on the environment (e.g. \citealt*{boselliandgavazzi_2014A&ARv..22...74B}).

Processes for that transformation include ram-pressure stripping
(\citealt*{gunnandgott.1972ApJ...176....1G,linandfaber.1983ApJ...266L..21L}) and galaxy harassment(\citealt{mooreetal.1998ApJ...495..139M}).
Ram-pressure stripping should be able to remove the galaxy's remaining gas from the system on short time
scales, so that star formation stops quickly. The effect of ram pressure stripping depends strongly on the
density of the environment and it is expected that their angular momentum and structure should be preserved
(\citealt{rysetal.2014MNRAS.439..284R}) while galaxy harassment by tidal interactions between a galaxy and the potential of the cluster can heat up the object, increasing the velocity dispersion, slow its rotation
down and remove stellar mass so that disks are transformed into more spheroidal objects
(\citealt{mooreetal.1998ApJ...495..139M}). In this case, a galaxy can lose some of its intrinsic angular
momentum. For a  more detailed review on these effects, see \citet*{boselliandgavazzi.2006PASP..118..517B,boselliandgavazzi_2014A&ARv..22...74B}. 

How exactly ram pressure stripping and harassment transform objects is still rather unclear.
\citet{rysetal.2013MNRAS.428.2980R} concluded that a transformation mechanism should be able to not only
lower the angular momentum but also increase the stellar concentration of dEs compared to their
presumed progenitors. \citet{tolobaetal.2015ApJ...799..172T} show that even a combination of these two
mechanisms can not easily remove all of the angular momentum, something which is needed to explain some
observations. Since ram pressure stripping is happening on short timescales, it might be a standard
mechanism to transform late type star-forming galaxies into dwarf early-type galaxies. After being in the
cluster for a long time, the galaxy goes through its center several times, during which it can heat up,
lose stellar rotation and also lose its disky structure. Since fast rotators in the outer parts of the
cluster are rotating faster than the fast rotators found in the inner part of the cluster
(\citealt{tolobaetal.2014ApJS..215...17T}, here after T14), \citet{springeletal.2005Natur.435..629S} suggested that
clusters were formed by the accretion of small groups of galaxies. According to this scenario, the
properties of slow and non-rotating dEs in the center of the cluster can be explained as well as the
existence of kinematically decoupled cores observed in some of the SMAKCED (Stellar content, MAss and Kinematics of Cluster Early-type Dwarfs) dEs
(\citealt{tolobaetal.2014ApJ...783..120T}) Therefore, galaxy clusters, as a place with many dwarf
ellipticals with a range of environmental properties, are excellent places to study the evolution and
formation of the dEs.

Not only can be used the morphology, or their kinematics, to study the evolution of galaxies: more detailed information can be obtained by studying the stellar populations, since the distribution of ages, 
metallicities and abundance ratios provide important information that can be used to study the evolutionary history of galaxies since chemical abundances of the gas are locked into the stars when they form. 
This information can be obtained via two general techniques. The first is by studying ages and abundances from observations of individual stars, which can be done for nearby galaxies where individual stars 
can be resolved. The second is by studying the integrated light from more distant galaxies to derive star formation histories and abundance distributions. This second technique 
is the only one currently 
available for galaxies at the distance of the Virgo cluster (16.5 Mpc, \citealt{meietal.2007ApJ...655..144M}). 

A non-trivial problem when analyzing the spectra of galaxies is the 
degeneracy between age and metallicity. One can break the age-metallicity degeneracy using a wide wavelength baseline, a combination of line indices, and accurate data. 
These then are compared to evolutionary stellar population models (e.g., \citealt{vazdekis.1999ApJ...513..224V}; \citealt*{bruzualandcharlot.2003MNRAS.344.1000B}; 
\citealt{thomasetal.2003MNRAS.343..279T,maraston.2005MNRAS.362..799M,schiavon.2007ApJS..171..146S,marastonetal.2009A&A...493..425M,vazdekisetal.2010MNRAS.404.1639V}; 
\citealt*{conroyandvandokkum.2012ApJ...747...69C}). By comparing model predictions with observational galaxy parameters age and metallicity distributions of the stars in 
that galaxy can be obtained. One can compare observations with models of a single age and metallicity, obtaining SSP-equivalent parameters. More complicated approaches, 
(e.g. STECKMAP (\citealt{ocvirketal.2006MNRAS.365...46O}), STARLIGHT (\citealt{cidfernandesetal.2005MNRAS.358..363C})) provide full star formation histories. A problem still 
remains the unicity of the solutions. As a rule of thumb one can say that the larger the wavelength range is, the more unique the solution.

Stellar population studies show that dEs have on average a lower metal content than giant ellipticals, as expected from the metallicity-luminosity 
relation (\citealt{michielsenetal.2008MNRAS.385.1374M,skillmanetal.1989ApJ...347..875S,sybliskaetal.2017MNRAS}). Their ages are somewhat younger on average. However, 
recent studies show that the stellar populations of dEs show indications of both young and old ages and a range in gradients (e.g. 
\citealt{kolevaetal.2009MNRAS.396.2133K,kolevaetal.2011MNRAS.417.1643K,denbroketal.2011MNRAS.414.3052D}; \citealt{rysetal.2015MNRAS.452.1888R}). 
Studies about detailed abundance ratios in dEs are scarce. \citet{gorgasetal.1997ApJ...481L..19G,michielsenetal.2008MNRAS.385.1374M} and 
\citet{sybliskaetal.2017MNRAS} show that [Mg/Fe] is similar to solar, lower than what is found in giant ellipticals. [Mg/Fe] can, however provide 
important information about the formation of a galaxy. Individual galaxy abundances are the result of chemical evolution, involving element enrichment in stars, 
supernova explosions and galactic winds from e.g. AGB stars. As a result, the measurement of abundances of many elements can give us a very detailed picture 
of the formation and evolution of a galaxy (see e.g. \citealt{tolstoyetal.2009ARA&A..47..371T} for Local group galaxies). We often call this way of studying 
galaxies galactic archaeology. 

Measured abundances of various elements allow us in principle to understand which enrichment processes have been dominant at different epochs of galaxy 
formation because of their different nucleosynthetic origin. It is thought that a group of lighter elements, the so called $\alpha$-elements, such as O and Mg, 
are produced by type II supernovae, supernovae originated from massive stars, which therefore occur on short timescales (\citealt*{wortheyetal.1992ApJ...398...69W}). 
Most of the Fe, on the other hand, is predominantly produced by a different group of supernovae, those of type Ia, which 
occur on a much longer timescale. As a result, elemental ratios of [$\alpha$/Fe] give us information about the relative contribution from the two types of supernovae at a 
given time, i.e., about the timescale of star formation. The observed correlation of [$\alpha$/Fe] abundance ratio and galaxy mass is an indication of the 
downsizing (\citealt*{vazdekisetal.2004ApJ...601L..33V}; \citealt{nelanetal.2005ApJ...632..137N,thomasetal.2005ApJ...621..673T}). For dEs, 
\citet{gorgasetal.1997ApJ...481L..19G}, measured that Virgo dEs are consistent with solar [$\alpha$/Fe] abundance ratio and showed that star formation must 
have happened on longer time scales in these systems. Several works confirmed these results and found that dEs have younger ages and lower metallicities 
than normal Es (\citealt{gehaetal.2003AJ....126.1794G,vanzeeetal.2004AJ....128.2797V}). Interpretation of abundance ratios of other elements is more complicated, 
and has been limited mostly to the Local Group. They are also used to obtain a more detailed picture and information on the IMF and 
SFH (\citealt{mcwilliam.1997ARA&A..35..503M}). In dEs, at the moment, 
very little information is available on abundance ratios of elements (apart from Mg and Fe), mainly because of the lack of high S/N spectra, but also because of 
the lack of methods to analyse them. Since this has changed in recent years, we have been able to start a program to obtain and analyze abundance ratios in dwarf ellipticals. 
The results on the pilot galaxy NGC 1396 are presented in \citet{mentzetal.2016MNRAS.463.2819M}. In the current paper a sample of 37 galaxies from the SMAKCED sample is analyzed.

To determine the abundance ratios from integrated spectra, we use the hybrid model calibration by \citet*{conroyetal.2014ApJ...780...33C}(CvD hereafter). 
They calculate spectra using standard stellar population models with solar abundance ratios, which are modified by using theoretical responses of spectra to 
abundance ratio variations, following a method developed by \citet{walcheretal.2009MNRAS.398L..44W}. Here, we will study the abundance ratios of a few elements in 
dwarf ellipticals, obtaining data which allow us to compare the formation history of dEs with those of giant ellipticals, the Milky Way, and other galaxies of the 
Local Group. We will focus on the Na doublet absorption features in the optical wavelength range at 5890 and 5896 \AA\ (NaD hereafter) and the Ca4227 line-strength 
indices, to study the abundances of Na and Ca, as well as the better studied Mg. Although interpreting the observational results is at present extremely difficult, 
already many conclusions can be derived by comparing them to other types of galaxies.

This work is part of the SMAKCED project, aimed at studying the nature of dEs in the Virgo Cluster. 
More details about the galaxies discussed here, and their properties, can be found in the other SMAKCED papers 
(\citealp{janzetal.2012ApJ...745L..24J,janzetal.2014ApJ...786..105J, tolobaetal.2014ApJ...783..120T, tolobaetal.2014ApJS..215...17T, tolobaetal.2015ApJ...799..172T}). 
In \citet{tolobaetal.2014ApJS..215...17T} a description is given of the spectroscopic part of the survey, while the H-band photometry is described in 
\citet{janzetal.2014ApJ...786..105J}. In \citet{tolobaetal.2014ApJS..215...17T}, kinematics of two dEs are presented that show kinematically decoupled cores. 
In \citet{tolobaetal.2015ApJ...799..172T} the stellar kinematics of dEs is presented as a function of projected distance to the center of the Virgo cluster.

The Virgo cluster is an ideal laboratory to study dEs because it contains hundreds of them, is close enough to resolve their detailed structure, and is a 
dynamically young cluster that is still evolving today \citep{binggellietal.1993A&AS...98..275B,bosellietal.2014A&A...570A..69B}. In this paper we focus on the 
abundance ratio distribution of dEs and compare them with other types of galaxies.

This paper is organized as follows. In Section 2, we present the general properties of our samples, the observations, and the main data reduction steps. 
We describe the measurements of age-sensitive and metallicity-sensitive Lick indices. In Section 3, we derive the ages and metallicity based on the Lick indices and 
the abundance ratios. In Section 4, our results are summarized and discussed. In Section 5, conclusions are given.

\section{Observations and Data Reduction}
Our sample consists of 37 galaxies, the full spectroscopic sample of the SMAKCED project. For each of them, we obtained data suitable for a 
detailed stellar population study, with relatively high spectral resolution and high signal-to-noise (S/N) ratio. Two galaxies in the sample (VCC1684, VCC2083) were not included, since no ages could be determined because of lack of observed Balmer lines.
The spectroscopic data were obtained at three different telescopes. Twenty six dEs were observed at the 4.2m WHT telescope using the double-arm ISIS spectrograph, 
of which the blue-arm covered the wavelength range 4200 - 5000 \AA\ and the red-arm covered the wavelength range 5500 - 6700 \AA\ . Ten dEs were observed at the 
2.5m INT telescope using the IDS spectrograph covering the wavelength range 4600 - 5600 \AA\ , while the remaining three dEs were observed at the 8m VLT  telescope 
using the FORS2 spectrograph that covers the wavelength range 4500 - 5600 \AA\ . 

Table~\ref{tab:properties} summarizes the main properties of these 37 dEs (see also T14).

The data were reduced following the standard procedure for long-slit spectra using the package REDUCEME (\citealp{cardiel.1999PhDT........12C}). 
Details on sample selection, observations and data reduction are presented in T14.

\begin{table*}
 \caption{Properties of the SMAKCED dEs. Column 1: galaxy name. Columns 2 and 3: right ascension and declination 
 in J2000. Columns 4 and 5: r-band magnitude (in the AB system) and half-light radius 
  \protect\citet{janzandlisker.2008ApJ...689L..25J,janzandlisker.2009ApJ...696L.102J}. Column 6: velocity dispersion.}
 \label{tab:properties}
 \begin{tabular}{cccccc} 
	    \hline
		Galaxy & RA  & Dec & M$_{r}$ & R$_{e}$ & $\sigma_\text{e}$ \\
		~      &   (J2000)        & (J2000)           & (mag)   & (arcsec)  &
		(km/s) \\
		\hline
		VCC0009	& 12:09:22.25 & 13:59:32.74 & -18.2 & 37.2  & 26.0$\pm3.9$\\
        VCC0021	& 12:10:23.15 & 10:11:19.04 & -17.1 & 15.2  & 28.9$\pm2.9$\\
        VCC0033	& 12:11:07.79 & 14:16:29.19 & -16.9 & 09.8  & 20.8$\pm4.9$\\
        VCC0170	& 12:15:56.34 & 14:26:00.33 & -17.6 & 31.3  & 26.6$\pm4.6$\\
        VCC0308	& 12:18:50.90 & 07:51:43.38 & -18.0 & 18.6  & 24.1$\pm2.4$\\
        VCC0389	& 12:20:03.29 & 14:57:41.70 & -18.1 & 18.0  & 30.9$\pm1.2$\\
        VCC0397	& 12:20:12.18 & 06:37:23.51 & -16.8 & 13.6  & 35.7$\pm1.9$\\
        VCC0437	& 12:20:48.10 & 17:29:16.00 & -18.0 & 29.5  & 40.9$\pm4.0$\\
        VCC0523	& 12:22:04.14 & 12:47:14.60 & -18.7 & 26.1  & 42.2$\pm1.0$\\
        VCC0543	& 12:22:19.54 & 14:45:38.59 & -17.8 & 23.6  & 35.1$\pm1.4$\\
        VCC0634	& 12:23:20.01 & 15:49:13.25 & -18.5 & 37.2  & 31.3$\pm1.6$\\
        VCC0750	& 12:24:49.58 & 06:45:34.49 & -17.0 & 19.5  & 43.5$\pm2.9$\\
        VCC0751	& 12:24:48.30 & 18:11:47.00 & -17.5 & 12.3  & 32.1$\pm2.4$\\
        VCC0781	& 12:25:15.17 & 12:42:52.59 & -17.2 & 13.4  & 38.0$\pm2.8$\\
        VCC0794	& 12:25:22.10 & 16:25:47.00 & -17.3 & 37.0  & 29.0$\pm3.9$\\
        VCC0856	& 12:25:57.93 & 10:03:13.54 & -17.8 & 16.5  & 31.3$\pm4.1$\\
        VCC0917	& 12:26:32.39 & 13:34:43.54 & -16.6 & 09.9  & 28.4$\pm1.4$\\
        VCC0940	& 12:26:47.07 & 12:27:14.17 & -17.4 & 19.8  & 40.4$\pm1.3$\\
        VCC0990	& 12:27:16.94 & 16:01:27.92 & -17.5 & 10.2  & 38.7$\pm1.3$\\
        VCC1010	& 12:27:27.39 & 12:17:25.09 & -18.4 & 22.2  & 44.6$\pm0.9$\\
        VCC1087	& 12:28:14.90 & 11:47:23.58 & -18.6 & 35.4  & 42.0$\pm1.5$\\
        VCC1122	& 12:28:41.71 & 12:54:57.08 & -17.2 & 17.3  & 32.1$\pm1.7$\\
        VCC1183	& 12:29:22.51 & 11:26:01.73 & -17.9 & 21.1  & 44.3$\pm2.4$\\
        VCC1261	& 12:30:10.32 & 10:46:46.51 & -18.5 & 23.8  & 44.8$\pm1.4$\\
        VCC1304	& 12:30:39.90 & 15:07:46.68 & -16.9 & 16.5  & 25.9$\pm2.7$\\
        VCC1355	& 12:31:20.21 & 14:06:54.93 & -17.6 & 30.3  & 20.3$\pm4.7$\\
        VCC1407	& 12:32:02.73 & 11:53:24.46 & -17.0 & 12.1  & 31.9$\pm2.1$\\
        VCC1431	& 12:32:23.41 & 11:15:46.94 & -17.8 & 09.8  & 52.4$\pm1.6$\\
        VCC1453	& 12:32:44.22 & 14:11:46.17 & -17.9 & 18.9  & 35.6$\pm1.4$\\
        VCC1528	& 12:33:51.61 & 13:19:21.03 & -17.5 & 09.6  & 47.0$\pm1.4$\\
        VCC1549	& 12:34:14.83 & 11:04:17.51 & -17.3 & 12.1  & 36.7$\pm2.3$\\
        VCC1695	& 12:36:54.85 & 12:31:11.93 & -17.7 & 24.0  & 24.4$\pm2.2$\\
        VCC1861	& 12:40:58.57 & 11:11:04.34 & -17.9 & 19.0  & 31.3$\pm1.5$\\
        VCC1895	& 12:41:51.97 & 09:24:10.28 & -17.0 & 16.3  & 23.8$\pm3.0$\\
        VCC1910	& 12:42:08.67 & 11:45:15.19 & -17.9 & 13.4  & 37.0$\pm1.2$\\
        VCC1912	& 12:42:09.07 & 12:35:47.93 & -17.9 & 22.5  & 36.0$\pm1.5$\\
        VCC1947	& 12:42:56.34 & 03:40:35.78 & -17.6 & 09.3  & 48.3$\pm1.3$\\
		\hline
  \end{tabular}
 \end{table*}

\subsection{Line-strength measurements}
\label{sec:maths} 

Observed spectral data can be studied either by fitting the full spectrum or by focusing on selected line indices. 
In this work we study selected line indices. We measured Lick indices (\citealp{wortheyetal.1994ApJS...94..687W}) in the LIS-5 \AA\ flux calibrated system 
(\citealp{vazdekisetal.2010MNRAS.404.1639V}). The new Line Index System (LIS) has numerous advantages over the Lick system. 
It is defined at three different resolutions, namely 5.0 \AA, 8.4 \AA, and 14.0 \AA. 
As such, it is particularly useful for analyzing small galaxies and globular clusters. 
As the resolution of the Lick/IDS library (FWHM $\sim$ 8-11 \AA) is much lower than what is available with our high resolution spectra, we broadened the spectra to the LIS-5 \AA\ system. 
The LIS-5 \AA\ system choice expresses a factor of 2 improvement in resolution over Lick/IDS system. Also, the fact that the LIS system is flux calibrated, 
makes it easier to reproduce data from other authors.

The spectra were broadened to the LIS-5 \AA\ system taking into account the velocity dispersion of the 
spectra. This ensures that every galaxy spectrum has the same spectral resolution, which is important and necessary to compare them with each other. 

We measured a total of 23 Lick indices (\citealp{faberetal.1985ApJS...57..711F, gorgasetal.1993ApJS...86..153G, wortheyetal.1994ApJS...94..687W}, 
\citealp*{wortheyandottaviani.1997ApJS..111..377W}). In this paper we use ${H}\gamma_\text{F}$, ${H}\beta$, Fe4383, Fe4531, Fe5270, Fe5335, Fe5406, 
Fe5709, Mgb, Ca4227 and NaD. The summary of the indices used are tabulated Table~\ref{tab:indicator}. However, because of the different wavelength range covered by our spectra, not all lines are used for all galaxies. Galaxies observed at the VLT do not cover the ${H}\gamma_\text{F}$, Fe4383 and NaD lines. Also, in some spectra which are obtained by WHT, ${H}\beta$ 
is located at the end of the spectrum, and since the spectra suffer from vignetting there, we could not determine this index here. In Table~\ref{tab:results_WHT} and 
Table~\ref{tab:results_INTandVLT}, we show the line-strength index measurements for each galaxy used.

\begin{table}
 \caption{A summary of the indices used}
 \label{tab:indicator}
 \begin{tabular}{ccc} 
	    \hline
Telescope	&	WHT	&	INT and VLT	\\
\hline
Age indicator	&	${H}\gamma_\text{F}$	&	 ${H}\beta $	\\
Metal indicator	&	Fe4383, Fe4531 	&	Fe4531, Fe5270, Fe5335 \\
 & Fe5709 & Fe 5406, Fe5709, Mgb \\
Na abundances	&	NaD	&	. . .	\\
Ca abundances	&	Ca4227	&	. . .	\\
Mg abundance	&	. . . 	&	Mgb	\\
\hline
  \end{tabular}
 \end{table}

\section{Results}
\subsection{Derived ages and metallicities}
Luminosity-weighted ages and metallicities are estimated using age-sensitive (${H}\beta$ and ${H}\gamma_\text{F}$) and metallicity-sensitive (Fe4383, Fe4531, Fe5270, 
Fe5335, Fe5406, Fe5709 and Mgb) Lick spectral indices (\citealp{worthey.1994ApJS...95..107W}) measured in the LIS-5 \AA\ system (\citealp{vazdekisetal.2010MNRAS.404.1639V}). 
For this we used an MCMC (Markov Chain Monte Carlo) code to derive the age and metallicity of the best fitting MILES single stellar population model.

We estimate the best luminosity weighted age and metallicity from all available index-index combinations by effectively computing the "distance" from our 
measured indices to all predicted values on the model grids, and finding the age and metallicity combination with the minimum total distance. 
The age and metallicity values for each index-index diagram and associated uncertainties were derived using 1000 MCMC iteration of the fit. 
To reduce the effects of the grid discretization, the two relevant parameters (e.g. age and metallicity) were interpolated. Uncertainties were calculated by 
performing Monte Carlo simulations, making use of the observational error in each index. In Table~\ref{tab:abundances} we list the best fitting parameters for 
ages and metallicities that are determined using combinations of all age sensitive lines and all metal indicators.

\begin{figure*}
\includegraphics[width=1\textwidth]{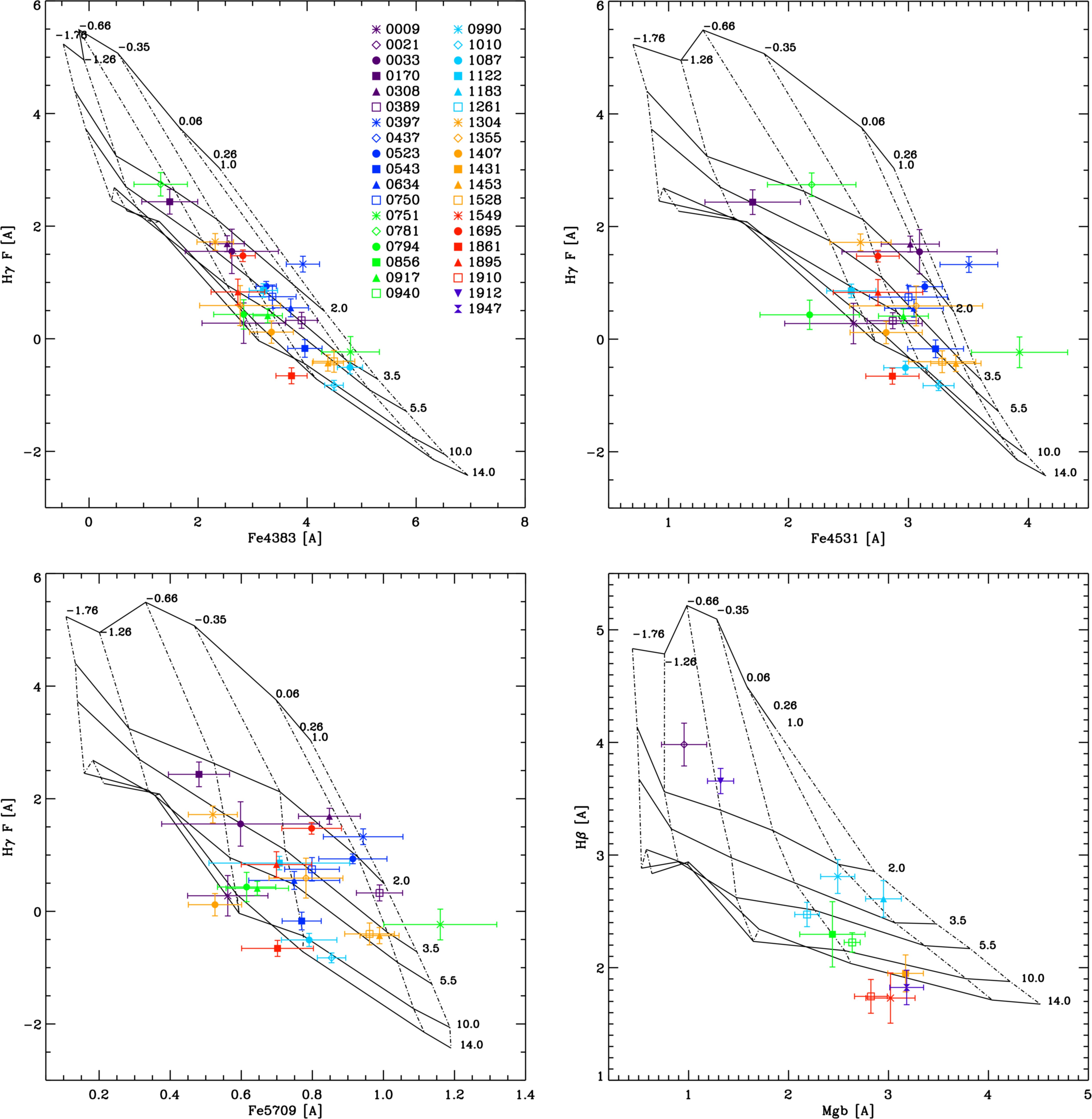}
\caption{Spectral index-index diagrams used to estimate the stellar populations using solar-scaled theoretical isochrone grids with IMF slope of 1.3 from 
Vazdekis et al. (2010) in the system LIS-5 \AA, solid lines indicate constant age 1.0, 2.0, 3.5, 5.5, 10.0 and 14.0 Gyr, respectively while dotted lines indicate 
constant [M/H] -1.76, -1.26, -0.65, -0.35, +0.06 and +0.26, respectively.}
\label{fig:index-index}
\end{figure*}

Figure~\ref{fig:index-index} shows index-index plots where we have restricted the age to the interval 1.0 - 14.0 Gyr, and the metallicity range from -1.76 to 0.26, which includes the range covered by the galaxies in our sample. We use the solar-scaled theoretical isochrones in the model grids from \citet{vazdekisetal.2010MNRAS.404.1639V} in Figure ~\ref{fig:index-index}. 
\begin{table*}
 \caption{Lick spectral indices measured at LIS-5 \AA\ resolution for WHT objects.}
 \label{tab:results_WHT}
 \begin{tabular}{ccccccc} 
	    \hline
		Galaxy & Ca4227 & ${H}\gamma_\text{F}$ & Fe4383 & Fe4531 & Fe5709 & NaD \\
		 & (\AA)  & (\AA) & (\AA) & (\AA) & (\AA) & (\AA)  \\
		\hline
VCC0033	&	1.25$\pm0.38$	    &	    1.55 $\pm0.39$	 &	 2.62$\pm0.85$       &       3.09$\pm0.65$	 &	 0.60$\pm0.22$   &	 0.83$\pm0.373$      \\
VCC0170	&	0.75$\pm0.21$	    &	    2.43 $\pm0.22$	 &	 1.48$\pm0.51$       &       1.70$\pm0.40$	 &	 0.48$\pm0.09$   &	 1.21$\pm0.173$      \\
VCC0308	&	0.99$\pm0.12$	    &	    1.69 $\pm0.14$	 &	 2.53$\pm0.32$       &       3.02$\pm0.24$	 &	 0.85$\pm0.09$   &	 1.57$\pm0.125$      \\
VCC0389	&	1.17$\pm0.14$	    &	    0.33 $\pm0.14$	 &	 3.90$\pm0.29$       &       2.87$\pm0.21$	 &	 0.99$\pm0.07$   &	 1.68$\pm0.095$      \\
VCC0397	&	1.11$\pm0.12$	    &	    1.33 $\pm0.14$	 &	 3.92$\pm0.31$       &       3.50$\pm0.24$	 &	 0.94$\pm0.11$   &	 1.92$\pm0.136$      \\
VCC0437	&	0.87$\pm0.25$    &	    -0.76$\pm0.27$	 &	 2.49$\pm0.57$       &       2.01$\pm0.41$	 &	 0.80$\pm0.10$   &	 2.08$\pm0.165$       \\
VCC0523	&	1.13$\pm0.07$	    &	    0.93 $\pm0.09$	 &	 3.25$\pm0.19$       &       3.14$\pm0.15$	 &	 0.91$\pm0.10$   &	 1.50$\pm0.177$       \\
VCC0543	&	1.41$\pm0.13$	    &	    -0.17$\pm0.16$	 &	 3.96$\pm0.32$       &       3.23$\pm0.23$	 &	 0.77$\pm0.06$   &	 1.95$\pm0.093$       \\
VCC0634 & 	1.21$\pm0.15$ 	& 0.55$\pm0.16$ & 3.70$\pm0.33$ & 3.05$\pm0.25$ & 0.75$\pm0.13$ & . . . \\
VCC0750	&	1.09$\pm0.20$	    &	    0.75 $\pm0.21$	 &	 3.36$\pm0.44$       &       3.00$\pm0.33$	 &	 0.80$\pm0.08$   &	 1.61$\pm0.122$       \\
VCC0751	&	1.62$\pm0.25$	    &	    -0.23$\pm0.27$	 &	 4.80$\pm0.53$       &       3.93$\pm0.40$	 &	 1.16$\pm0.16$   &	 . . .	              \\
VCC0781	&	0.66$\pm0.22$	    &	    2.74 $\pm0.21$	 &	 1.31$\pm0.49$       &       2.19$\pm0.37$	 &	 . . . 	         &	 . . .	              \\
VCC0794	&	1.06$\pm0.24$	    &	    0.43 $\pm0.26$	 &	 2.84$\pm0.55$       &       2.18$\pm0.41$	 &	 0.62$\pm0.08$   &	 1.45$\pm0.117$      \\
VCC0917	&	0.99$\pm0.10$	    &	    0.41 $\pm0.13$	 &	 3.28$\pm0.27$       &       2.96$\pm0.21$	 &	 0.65$\pm0.09$   &	 1.26$\pm0.129$       \\
VCC1010	&	1.42$\pm0.07$	    &	    -0.83$\pm0.09$	 &	 4.49$\pm0.17$       &       3.25$\pm0.13$	 &	 0.85$\pm0.04$   &	 2.33$\pm0.054$       \\
VCC1087	&	1.33$\pm0.10$	    &	    -0.51$\pm0.12$	 &	 4.79$\pm0.23$       &       2.98$\pm0.18$	 &	 0.79$\pm0.08$   &	 . . .	             \\
VCC1122	&	1.11$\pm0.10$	    &	    0.86 $\pm0.12$	 &	 3.20$\pm0.26$       &       2.52$\pm0.21$	 &	 0.71$\pm0.20$   &	 . . .	              \\
VCC1304	&	0.81$\pm0.13$	    &	    1.72 $\pm0.15$	 &	 2.31$\pm0.34$       &       2.60$\pm0.26$	 &	 0.52$\pm0.07$   &	 1.95$\pm0.097$       \\
VCC1355	&	1.39$\pm0.33$	    &	    0.59 $\pm0.35$	 &	 2.77$\pm0.74$       &       3.06$\pm0.55$	 &	 0.78$\pm0.10$   &	 1.32$\pm0.188$       \\
VCC1407	&	0.83$\pm0.18$	    &	    0.12 $\pm0.20$	 &	 3.35$\pm0.40$       &       2.81$\pm0.30$	 &	 0.53$\pm0.08$   &	 1.55$\pm0.129$       \\
VCC1453	&	1.52$\pm0.13$	    &	    -0.43$\pm0.14$	 &	 4.38$\pm0.29$       &       3.40$\pm0.21$	 &	 0.99$\pm0.06$   &	 2.08$\pm0.065$       \\
VCC1528	&	1.27$\pm0.18$	    &	    -0.40$\pm0.19$	 &	 4.49$\pm0.38$       &       3.28$\pm0.28$	 &	 0.96$\pm0.07$   &	 2.33$\pm0.093$       \\
VCC1695	&	1.03$\pm0.09$	    &	    1.47 $\pm0.10$	 &	 2.82$\pm0.23$       &       2.75$\pm0.18$	 &	 0.80$\pm0.08$   &	 1.63$\pm0.123$       \\
VCC1861	&	1.43$\pm0.11$	    &	    -0.66$\pm0.14$	 &	 3.71$\pm0.29$       &       2.87$\pm0.22$	 &	 0.70$\pm0.10$   &	 . . .	              \\
VCC1895	&	1.09$\pm0.20$	    &	    0.83 $\pm0.23$	 &	 2.73$\pm0.49$       &       2.75$\pm0.37$	 &	 0.70$\pm0.10$   &	 1.26$\pm0.155$       \\

		\hline
  \end{tabular}
 \end{table*}
 
 \begin{table*}
 \caption{Lick spectral indices measured at LIS-5 \AA\ resolution for INT and VLT objects.}
 \label{tab:results_INTandVLT}
 \begin{tabular}{cccccccc} 
	    \hline
		Galaxy & Fe4531 & ${H}\beta $ & Mgb & Fe5270 & Fe5335 & Fe5406 & Fe5709\\
		 & (\AA)  & (\AA) & (\AA) & (\AA) & (\AA) & (\AA) & (\AA) \\
		\hline
VCC0021	&	1.23$\pm0.40$	    &	    3.98$\pm0.19$	&	0.96$\pm0.23$	    &	    1.25$\pm0.25$	&	1.52$\pm0.29$	    &	    0.51$\pm0.23$	&	0.11$\pm0.20$\\
VCC0856	&	3.35$\pm0.62$	    &	    2.30$\pm0.29$	&	2.44$\pm0.33$	    &	    2.38$\pm0.36$	&	1.87$\pm0.41$	    &	    0.67$\pm0.32$	&	1.39$\pm0.27$\\
VCC0940	&	2.33$\pm0.17$	    &	    2.22$\pm0.09$	&	2.64$\pm0.08$	    &	    2.45$\pm0.09$	&	1.78$\pm0.10$	    &	    1.30$\pm0.07$	&	...\\
VCC0990	&	3.20$\pm0.30$	    &	    2.81$\pm0.15$	&	2.49$\pm0.17$	    &	    2.62$\pm0.19$	&	2.29$\pm0.21$	    &	    1.58$\pm0.16$	&	0.83$\pm0.15$\\
VCC1183	&	3.49$\pm0.35$	    &	    2.61$\pm0.16$	&	2.95$\pm0.18$	    &	    2.82$\pm0.20$	&	2.23$\pm0.23$	    &	    1.59$\pm0.17$	&	0.95$\pm0.15$\\
VCC1261	&	2.33$\pm0.21$	    &	    2.47$\pm0.11$	&	2.19$\pm0.12$	    &	    2.52$\pm0.13$	&	2.13$\pm0.15$	    &	    1.47$\pm0.12$	&	0.94$\pm0.11$\\
VCC1431	&	3.53$\pm0.33$	    &	    1.95$\pm0.16$	&	3.17$\pm0.18$	    &	    2.46$\pm0.19$	&	1.93$\pm0.22$	    &	    1.41$\pm0.17$	&	0.46$\pm0.15$\\
VCC1549	&	2.72$\pm0.46$	    &	    1.73$\pm0.22$	&	3.02$\pm0.25$	    &	    2.85$\pm0.26$	&	2.48$\pm0.30$	    &	    1.74$\pm0.22$	&	0.99$\pm0.20$\\
VCC1910	&	3.24$\pm0.33$	    &	    1.75$\pm0.15$	&	2.82$\pm0.16$	    &	    2.50$\pm0.18$	&	2.84$\pm0.19$	    &	    2.06$\pm0.15$	&	0.30$\pm0.13$\\
VCC1912	&	2.31$\pm0.24$	    &	    3.66$\pm0.11$	&	1.32$\pm0.13$	    &	    2.21$\pm0.14$	&	2.30$\pm0.16$	    &	    1.15$\pm0.13$	&	0.16$\pm0.11$\\
VCC1947	&	3.26$\pm0.32$	    &	    1.82$\pm0.15$	&	3.18$\pm0.17$	    &	    2.99$\pm0.18$	&	2.69$\pm0.20$	    &	    1.85$\pm0.16$	&	0.94$\pm0.14$\\

\hline
  \end{tabular}
 \end{table*}


\subsection{Elemental abundance ratios}
{To calculate the abundance ratios [E/Fe] from the indices we first calculate age and metallicity as described above, assuming that the galaxies can be represented by an SSP model. 
We then measure the difference between the observed index related to the element E and the index value expected from stellar population models for the age and metallicity that we measure for the galaxy, and divide this 
difference by the sensitivity of the index to variations in [E/Fe] (equation 1).  For Na (NaD) and Mg (Mgb) we use 
the Na-MILES model of (\citealt{labarberaetal.2017MNRAS.464.3597L})  based on Teramo isochrones, with variable Mg and Na abundance ratios, and with a (standard) bimodal IMF with $\Gamma_b$=1.3.
Unfortunately, Mgb was measured only in a few galaxies, those observed with the VLT and the INT. For Ca, we do the same using the Ca4227 index.  The problem here is that for this element no stellar population models are available. The only available model, by CvD, has solar metallicity, and a fixed, old, age of 13.5 Gyr, which is different from the ages and metallicities of the objects we discuss here. In this paper we will use this latter model, but move the analysis and discussion to Appendix A, since we do not know how appropriate this is.

So, for element {\it E} based on index {\it i} the elemental abundance ratios were calculated using 
\begin{equation}
    [E_{i}/Fe]=\frac{i_{observed} - i_{model} }{\frac{\Delta{i_{model}}}{\Delta{[E_{i}/Fe]_{model}}}}.
	\label{eq:quadratic}
\end{equation}
where, $\it{i_{observed}}$ is the value of the index measured from the observations, $\it{i_{model}}$ is the index expected from the model and $\it{E_{i}}$ 
is the elemental abundance.

In Table~\ref{tab:abundances}, we provide the elemental abundances for each galaxy determined according to the procedure described above.
The spectral range covered allowed the measurement of [Ca/Fe] and [Na/Fe] for most galaxies that are observed by WHT, and [Mg/Fe] for some that are observed by INT and VLT. 
We do not have galaxies for which all three abundance ratios have been measured.
 
To establish the reliability of these abundance ratios,  we made a comparison of the sensitivity of $\Delta{NaD}$\ to
variations in [Na/Fe] and $\Delta{Mgb}$\ to variations in [Mg/Fe] between the SSP models of CvD and MILES. For a meaningful comparison, we have computed some additional $\alpha$ enhancement SSP models 
for a total metallicity of [M/H]=0.3 (for [Fe/H]~0.0, from equation 4 of \citealt{vazdekisetal.2012MNRAS.424..157V}). 
For the Na comparison, a Na-MILES model (\citealt{labarberaetal.2017MNRAS.464.3597L}) is computed for the same age and metallicity as CvD.

We find that the Na-MILES model give a slightly larger sensitivity for Na than the one of CvD and that we get similar results from both CvD and MILES for Mgb. 
Note, however, that the sensitivities vary considerably when going to lower ages and metallicities.

We report the values of the $\Delta{NaD}$\ and $\Delta{Mgb}$\ variations as a function of these elemental abundance ratios for models with different ages and metallicities 
in Table~\ref{tab:comparision}. When we compare the sensitivities at the same age, we find that the NaD sensitivity generally increases with metallicity, while the one for Mgb generally decreases. When one increases the age, both sensitivities tend to increase.

\begin{figure*}
	\includegraphics[width=.95\textwidth]{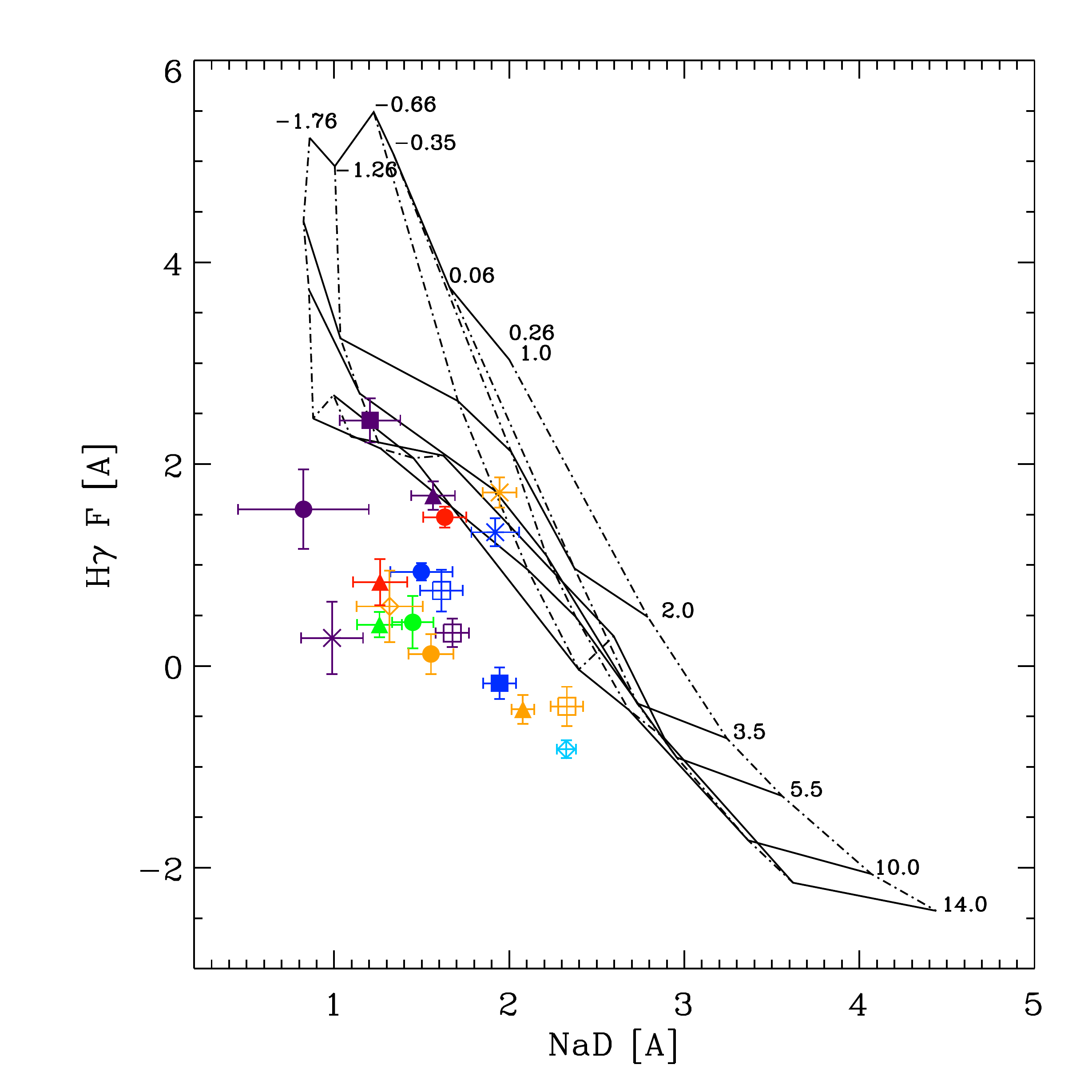}
    \caption{Example of a spectral index-index diagram for ${H}\gamma_\text{F}$ versus NaD, showing that the NaD values are much lower than the predicted by models. Models from Vazdekis et al. 2010. Constant age and [M/H] as in Figure~\ref{fig:index-index}.}
    \label{fig:index_index_1aug_hgf_nad}
\end{figure*}

\begin{figure*}
	\includegraphics[width=.93\textwidth]{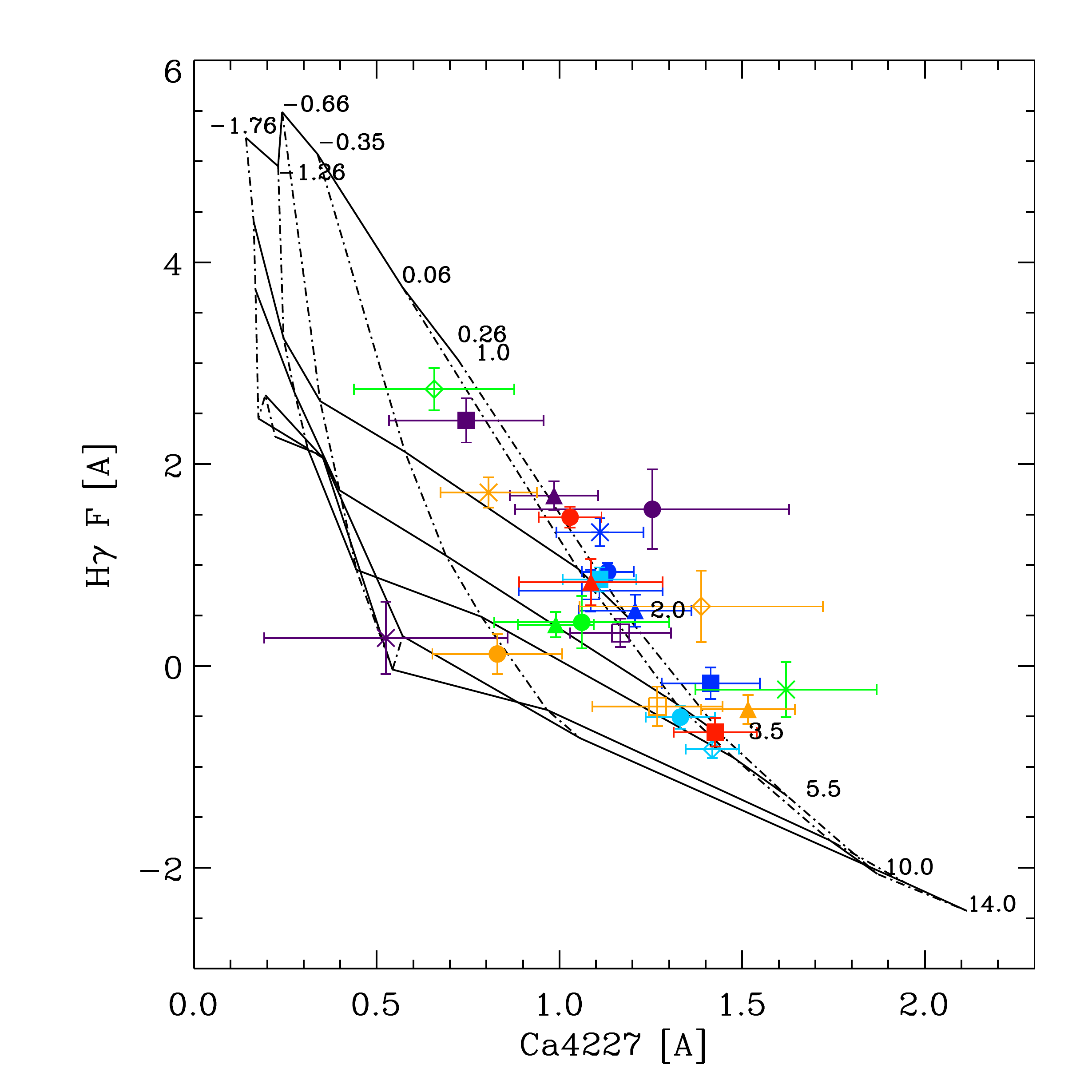}
    \caption{Example of spectral index-index diagram for ${H}\gamma_\text{F}$ versus Ca4227, showing that the Ca4227 values are slightly higher than predicted by models. Models from Vazdekis et al. 2010. Constant age and [M/H] as in Figure~\ref{fig:index-index}.}
    \label{fig:index_index_1aug_hgf_ca4227}
\end{figure*}

\begin{figure*}
	\includegraphics[width=0.95\textwidth]{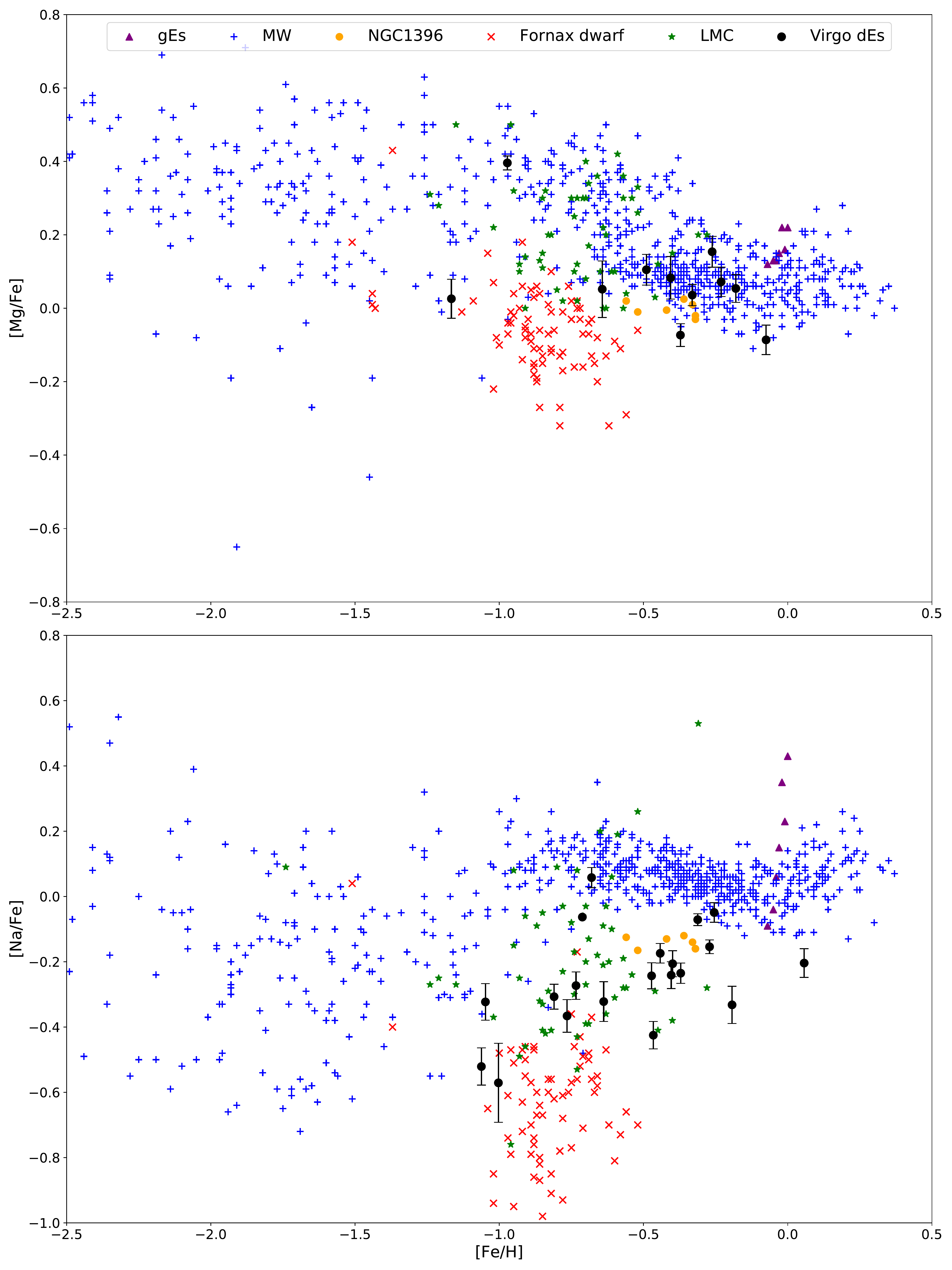}
    \caption{Mg and Na abundances as a function of metallicity [Fe/H]. Blue pluses are the Milky Way stars from Venn et al. (2004), red crosses are from the Fornax local dwarf from \citet{Letarteetal.2010A&A...523A..17L} and \citet{2003AJ....125..684S}, green asterisks are the LMC red giants from Pomp\'eia et al. (2008), purple triangles are giant ellipticals from Conroy et al. (2014), orange circles are various radial bins of the NGC1396 in the Fornax cluster from Metz et al. (2016) and black circles are the dEs in Virgo cluster analyzed here.}
    \label{fig:abundances_error}    
\end{figure*}

\begin{table*}
 \caption{Elemental abundances, metallicity and ages. Column 1: galaxy name. Column 2 and 3: measurement of [Ca/Fe] and [Na/Fe] for galaxies that are observed with the WHT. Column 4: measurement of [Mg/Fe] for galaxies that are observed using the INT and VLT. Column 5 and 6: metallicity [M/H] and logarithmic ages in Gyr with errors.}
 \label{tab:abundances}
 \begin{tabular}{cccccc} 
	    \hline
		Galaxy & [Ca/Fe] & [Na/Fe] & [Mg/Fe] & [Fe/H] & log (age) \\
		 & & & & & (Gyr)\\
        \hline
VCC0009	&	0.05	&	-0.52	&	. . . 	&	-1.06	$\pm	0.15	$	&	0.94	$\pm	0.03	$	\\
VCC0021	&	. . . 	&	. . . 	&	0.03	&	-1.17	$\pm	0.18	$	&	0.20	$\pm	0.07	$	\\
VCC0033	&	0.32	&	-0.57	&	. . . 	&	-1.00	$\pm	0.01	$	&	0.83	$\pm	0.17	$	\\
VCC0170	&	0.17	&	-0.32	&	. . . 	&	-1.05	$\pm	0.08	$	&	0.47	$\pm	0.09	$	\\
VCC0308	&	0.14	&	-0.24	&	. . . 	&	-0.40	$\pm	0.24	$	&	0.34	$\pm	0.04	$	\\
VCC0389	&	0.15	&	-0.23	&	. . . 	&	-0.37	$\pm	0.16	$	&	0.70	$\pm	0.04	$	\\
VCC0397	&	0.05	&	-0.20	&	. . . 	&	0.06	$\pm	0.35	$	&	0.23	$\pm	0.04	$	\\
VCC0437	&	0.12	&	-0.06	&	. . . 	&	-0.71	$\pm	0.01	$	&	0.98	$\pm	0.07	$	\\
VCC0523	&	0.17	&	-0.33	&	. . . 	&	-0.19	$\pm	0.29	$	&	0.41	$\pm	0.03	$	\\
VCC0543	&	0.21	&	-0.17	&	. . . 	&	-0.44	$\pm	0.07	$	&	0.83	$\pm	0.05	$	\\
VCC0634	&	0.30	&	. . . 	&	. . . 	&	-0.93	$\pm	0.35	$	&	0.84	$\pm	0.30	$	\\
VCC0750	&	0.13	&	-0.24	&	. . . 	&	-0.47	$\pm	0.10	$	&	0.62	$\pm	0.06	$	\\
VCC0751	&	0.45	&	. . . 	&	. . . 	&	-0.93	$\pm	0.36	$	&	0.85	$\pm	0.29	$	\\
VCC0781	&	0.11	&	. . . 	&	. . . 	&	-0.90	$\pm	0.24	$	&	0.84	$\pm	0.30	$	\\
VCC0794	&	0.25	&	-0.31	&	. . . 	&	-0.81	$\pm	0.03	$	&	0.90	$\pm	0.08	$	\\
VCC0856	&	. . . 	&	. . . 	&	0.05	&	-0.64	$\pm	0.31	$	&	0.92	$\pm	0.16	$	\\
VCC0917	&	0.08	&	-0.43	&	. . . 	&	-0.47	$\pm	0.12	$	&	0.69	$\pm	0.05	$	\\
VCC0940	&	. . . 	&	. . . 	&	0.40	&	-0.97	$\pm	0.35	$	&	0.84	$\pm	0.29	$	\\
VCC0990	&	. . . 	&	. . . 	&	0.07	&	-0.23	$\pm	0.12	$	&	0.43	$\pm	0.08	$	\\
VCC1010	&	0.18	&	-0.07	&	. . . 	&	-0.31	$\pm	0.10	$	&	0.94	$\pm	0.03	$	\\
VCC1087	&	0.35	&	. . . 	&	. . . 	&	-0.94	$\pm	0.25	$	&	0.84	$\pm	0.30	$	\\
VCC1122	&	0.27	&	. . . 	&	. . . 	&	-0.93	$\pm	0.30	$	&	0.85	$\pm	0.29	$	\\
VCC1183	&	. . . 	&	. . . 	&	0.15	&	-0.26	$\pm	0.02	$	&	0.67	$\pm	0.11	$	\\
VCC1261	&	. . . 	&	. . . 	&	0.04	&	-0.33	$\pm	0.13	$	&	0.52	$\pm	0.06	$	\\
VCC1304	&	0.15	&	0.06	&	. . . 	&	-0.68	$\pm	0.10	$	&	0.55	$\pm	0.07	$	\\
VCC1355	&	0.34	&	-0.32	&	. . . 	&	-0.64	$\pm	0.01	$	&	0.75	$\pm	0.12	$	\\
VCC1407	&	0.11	&	-0.27	&	. . . 	&	-0.73	$\pm	0.07	$	&	0.95	$\pm	0.05	$	\\
VCC1431	&	. . . 	&	. . . 	&	0.11	&	-0.49	$\pm	0.02	$	&	1.07	$\pm	0.06	$	\\
VCC1453	&	0.23	&	-0.15	&	. . . 	&	-0.27	$\pm	0.13	$	&	0.76	$\pm	0.05	$	\\
VCC1528	&	0.14	&	-0.05	&	. . . 	&	-0.25	$\pm	0.05	$	&	0.79	$\pm	0.06	$	\\
VCC1549	&	. . . 	&	. . . 	&	0.08	&	-0.41	$\pm	0.09	$	&	1.04	$\pm	0.09	$	\\
VCC1695	&	0.15	&	-0.21	&	. . . 	&	-0.40	$\pm	0.22	$	&	0.39	$\pm	0.05	$	\\
VCC1861	&	0.38	&	. . . 	&	. . . 	&	-0.90	$\pm	0.57	$	&	0.85	$\pm	0.29	$	\\
VCC1895	&	0.22	&	-0.37	&	. . . 	&	-0.77	$\pm	0.15	$	&	0.82	$\pm	0.11	$	\\
VCC1910	&	. . . 	&	. . . 	&	0.05	&	-0.37	$\pm	0.40	$	&	0.93	$\pm	0.11	$	\\
VCC1912	&	. . . 	&	. . . 	&	-0.07	&	-0.07	$\pm	0.28	$	&	0.14	$\pm	0.01	$	\\
VCC1947	&	. . . 	&	. . . 	&	-0.09	&	-0.93	$\pm	0.47	$	&	0.88	$\pm	0.11	$	\\
        
 	\hline
  \end{tabular}
 \end{table*}
 
\begin{table}[H]
 \caption{Comparison of MILES and CvD model predictions (column 1) with varying metallicity (column 2) and age (column 3). Columns 4 and 5: $\Delta{NaD}$ variations in [Na/Fe], $\Delta{Ca4227}$ variations in [Ca/Fe] and $\Delta{Mgb}$ variations in [Mg/Fe] based on the models shown in col.1}
 \label{tab:comparision}
 \begin{tabular}{cccccc} 
	    \hline
		Model & [M/H] & Age  & $\frac{\Delta{NaD_{*}}}{\Delta{[Na/Fe]_{*}}} $  & $\frac{\Delta{Mgb_{*}}}{\Delta{[Mg/Fe]_{*}}} $ &$\frac{\Delta{Ca4227_{*}}}{\Delta{[Ca/Fe]_{*}}} $ \\
		(*) & dex & (Gyr) & & &\\
		\hline
CvD	&	0.00	&	13.50	&	3.085	&	4.260 & 2.790	\\
MILES	&	0.06	&	14.00	&	3.547	&	4.350& ...	\\
\hline
MILES	&	-0.35	&	1.00	&	0.973	&	3.133	& ...\\
MILES	&	-0.35	&	7.00	&	2.317	&	4.830	&	...	\\
MILES	&	-0.35	&	14.00	&	2.683	&	5.609	&	...	\\
MILES	&	0.06	&	1.00	&	1.453	&	3.009	&	...	\\
MILES	&	0.06	&	7.00	&	3.167	&	3.298	&	...	\\
MILES	&	0.26	&	1.00	&	1.583	&	3.463	&	...	\\
MILES	&	0.26	&	7.00	&	3.380	&	3.336	&	...	\\
MILES	&	0.26	&	14.00	&	3.830	&	3.768	&	...	\\
\hline
  \end{tabular}
 \end{table}

\section{Discussion}
We present ages, metallicities and abundance ratios obtained for 37 dEs within an aperture size of R$_e$/8. This aperture size is commonly used in conventional long-slit studies. 
The radius has been chosen to provide a measure of the stellar populations in the central regions of these galaxies, in a region with a constant relative size.  Following the classification of  
\citet{liskeretal.2006AJ....132..497L,liskeretal.2006AJ....132.2432L,liskeretal.2007ApJ...660.1186L} using high-pass filtered Sloan Digital Sky Survey (SDSS; 
\citealt{2006ApJS..162...38A}), 36 of the galaxies of our sample are classified as nucleated. VCC0021 and VCC1431 are galaxies with large blue core. All the analysis comes from the central regions, 
in which the nuclear cluster's light is contributing  significantly. The resulting abundance
ratios can thus provide insight about the luminosity-weighted stellar population in this region. Although we could have tried to derive two-burst or more complicated Star Formation Histories, 
it would have been very hard with the current low-resolution data to be able to distinguish between a one- or two-burst solution (see e.g. \citealt{rysetal.2015MNRAS.452.1888R,mentzetal.2016MNRAS.463.2819M}).
For that reason, we postpone this to a future paper, in which we analyze spectra at a spectral resolution of R=5000 using the SAMI instrument at the AAT.

One of the main results of this paper is the unusual behaviour of the Na abundances in dEs, when compared with massive ellipticals and Local Group dwarfs. 
For a sample of quiescent dwarfs with effective velocity dispersion in the range 20 - 55\,km\,s$^{-1}$ and with absolute r-band magnitude ranging from -19 to -16, we find that our sample of dEs is underabundant in 
Na when compared to the solar neighbourhood (Figure~\ref{fig:abundances_error}). At the same time, the Mg abundances are around solar. In the following, we try to argue what this means for the evolution of dwarf ellipticals, 
based on what we know from theory, and observations of individual stars in the Milky Way and the Local Group, and integrated abundance ratios in giant elliptical galaxies. For dwarf ellipticals, such an 
analysis using integrated light has not been done before, although several papers have tried to derive abundance ratios in giant ellipticals 
(e.g. \citealt{worthey.1998PASP..110..888W, thomasetal.2010MNRAS.404.1775T,wortheyetal.2011ApJ...729..148W,conroyetal.2014ApJ...780...33C,spinielloetal.2014MNRAS.438.1483S,smithetal.2015MNRAS.454L..71S,yamadaetal.2006ApJ...637..200Y}).

In this paper we are analyzing light elements. It is our current belief that in the most-studied environments the relative abundance [X/Fe] of an element X at low [Fe/ H] represents the abundance ratio 
from only those sources of material processed through the nucleosynthesis channels that were active at very early times, i.e. SNII from core collapse of massive stars, events characteristic of the earliest epochs 
of star formation \citep*{cohenandhuang.2009ApJ...701.1053C}. After these early times, [X/Fe] is modified by other sources, such as  SNIa, SNII, AGB stars, novae, etc., and also by the accretion of primordial 
material and of galactic winds. This gives a knee-like pattern in the relation between [X/Fe] and [Fe/ H], of which the position of the bend is determined by the delay time between the core collapse SNII and 
the other processes. This delay depends on various parameters (IMF, star formation efficiency, the rate of mass loss in stars, and the rate of accretion of primordial gas consisting mostly of H), as well as 
on the element production yields (e.g. \citealt{greggioetal.2008MNRAS.388..829G}). These processes lead to the characteristic knee in the Milky Way, where low metallicity halo stars generally have [Mg/Fe] 
values around 0.4, a value reached by SNII enrichment only, going down to solar mass disk stars, with [Mg/Fe] values around solar, an equilibrium values reached when all processes have contributed. From these 
solar abundance ratios in the disk of our Milky Way, one can then conclude that star formation has been taken place on long timescales compared to the halo.

We will now have a more detailed look at the individual abundance ratios.

\subsection{Na abundances}
Na is believed to be made in the interiors of massive stars and to depend on the neutron excess, which in turn depends on the initial heavy element abundance in the star. 
Na thus has both a primary and a secondary nucleosynthesis channel (\citealt{arnett.1971ApJ...166..153A,clayton.2003Ap&SS.285..353C}). Ni is assumed to originate predominantly from SNe Ia. However, 
the production of Ni might also be linked to the production of Na in SNe II (\citealt{thielemannetal.1990ApJ...349..222T,timmesetal.1995ApJS...98..617T}). The amount of Na produced is controlled by the 
neutron excess, where ${}^{23}Na$ is the only stable neutron rich isotope produced in significant quantity during the C and O burning stage. The production of Ni depends on the neutron excess and the neutron 
excess will depend primarily on the amount of ${}^{23}Na$ previously produced. Hence, a Na-Ni correlation is expected when the chemical enrichment is dominated by SNe II. The advent of the SNe Ia explosions 
can break (or flatten) this relationship, as Ni is produced without Na in the standard model of SN Ia (e.g., \citealt{Iwamotoetal.1999ApJS..125..439I}). Since the neutron excess is strongly metallicity-dependent, 
this could explain the low [Na/Fe] that are being found for the Fornax dwarf (\citealt{Letarteetal.2010A&A...523A..17L}). At the same time, it would explain the high values in giant ellipticals (see later).

It is important to mention is that the abundance ratios found here are very different from those in other stellar systems, most massive Galactic globular clusters, and giant elliptical galaxies. In most massive 
globular clusters a strong  Na-O anticorrelation is observed in the RGB stars (\citealt{kraft.1994PASP..106..553K}, and see reviews of \citealt{carretta.2016arXiv161104728C,grattonetal.2001A&A...369...87G}). 
For these stars, Oxygen is depleted, and Na enhanced, just like N. Since this effect is not see in the interior of the presently observed GC low mass stars, it is thought that this is a second generation 
enrichment effect from massive stars, who have enhanced their Na-abundance during C-burning through the NeNa cycle (\citealt{langeretal.1993PASP..105..301L}).  It is thought that second generation (SG) 
stars are formed by nuclear ejecta processed in the most massive first generation (FG) stars, diluted with different amounts of unprocessed gas, generating a number of anticorrelations, including the 
one of Na and O. This process, as far as we know, does not take place in the halo of our Milky Way, and in field stars of local group galaxies. In this paper we see that this is also not the case in 
field stars of dwarf ellipticals in nearby galaxy clusters.

Four strong Na absorption features can be found in the optical and NIR wavelength range: NaD (5890 and 5896 \AA), NaI 0.82 (8183 and 8195 \AA), NaI1.14 (11400 \AA) and NaI2.21 (22100 \AA). 
The equivalent widths(EWs) of NaD were studied first by \citet{oconnel.1976ApJ...206..370O} and \citet{peterson.1976ApJ...210L.123P}, where they reported that NaD was much stronger than expected from 
Ca and Fe indices in giant early-type galaxies.
\citet{spinielloetal.2015ApJ...803...87S} reported that the NaD feature is very sensitive to [Na/Fe] variations, while the NaI index seems to depend mainly on the IMF (e.g. \citealt{vazdekisetal.2012MNRAS.424..157V}).
\citet{labarberaetal.2017MNRAS.464.3597L} 
perform detailed fits to all 4 Na-lines in a number of giant ellipticals, and find that both [Na/Fe] needs to be considerably larger than solar, and the IMF-slope needs to be dwarf-enhanced (\citealt{smithetal.2015MNRAS.454L..71S}). 
For dwarf ellipticals, however, there are no indications that the IMF-slope is different from our Galaxy (\citealt{mentzetal.2016MNRAS.463.2819M}). Here, we find the behaviour that [Na/Fe] is lower than solar, 
opposite to the behavior in giant ellipticals.

Remarkable is the strong trend between [Na/Fe] and [Fe/H], already mentioned by \citet{mentzetal.2016MNRAS.463.2819M} when joining the Fornax dwarf with our dwarf ellipticals and the giant ellipticals. 
For [Fe/H] values of $\sim$ -0.8 very low [Na/Fe] values are obtained of $\sim$ -0.6 - -0.8 in the Fornax dwarf. This contrasts with the high, positive [Na/Fe] values that are found in giant ellipticals. Such a 
strong correlation is what you could expect if the Na-abundance depends strongly on the neutron excess, or equivalently, the metallicity.

\subsection{Abundance ratios and the formation of dEs}

In this paper we find [Mg/Fe] values that are very close to solar (the mean value of the [Mg/Fe] is 0.07), or slightly larger. [Na/Fe] is (the mean value of the [Na/Fe] is -0.25), however, considerably lower than solar. 

When comparing to stellar populations that show similar abundance patterns, there are not many, but one could consider the younger, and more metal-rich stellar populations in the center of the 
Fornax Dwarf galaxy, published by \citet{Letarteetal.2010A&A...523A..17L}. The majority of those stars are 1-4 Gyr old, and  have unusually low [Mg/Fe], [Ca/Fe] and [Na/Fe] compared to the Milky 
Way stellar populations at the same [Fe/H], and are therefore at the end of their chemical evolution. The difficulty is that, although [Mg/Fe] is approximately solar for these stars, [Ca/Fe] lies 
considerably below this value. \citet{Letarteetal.2010A&A...523A..17L} hypothesize that this means that a large fraction of Ca and Ti is produced in processes that do not produce much Mg, such in 
SNe Ia. In this way, the low [Ca/Fe] and [Ti/Fe] could be a consequence of the low-metallicity of their progenitors compared to the Milky Way. At the same time, [Na/Fe] is found to be rather low 
(between -0.9 and -0.4), but correlating strongly with [Ni/Fe] \citep*{nissenandschuster.1997A&A...326..751N,nissenandschuster.2009IAUS..254..103N}.

For the dwarf ellipticals in Virgo analyzed in this paper, we conclude, analogous to \citet{Letarteetal.2010A&A...523A..17L} that the low [Mg/Fe] values (w.r.t. the thick disk of the Milky Way and giant ellipticals) show that the galaxies have undergone 
a considerable amount of chemical evolution. This means that the galaxies are not uniformly old, but have extended star formation histories, similar to many of the Local Group galaxies.  [Na/Fe] is lower than solar, but still higher than in the Fornax dwarf. This is expectable, 
since the metallicities of the dwarf ellipticals in Virgo are a bit larger, resulting in a larger neutron excess and higher Na-abundances.

For dEs we find that (a) [Mg/Fe] $\sim$ 0 and (b) [Na/Fe] < 0.  Result (a) implies that star formation is slow, like in the Milky Way disk. 
Result (b) is consistent with the same formation mechanism. Just like the stars of \citet{Letarteetal.2010A&A...523A..17L} in the center of the Fornax dwarf galaxy, 
the stars in the dEs have undergone a considerable amount of enrichment and have an extended star formation history. The dependence of [Na/Fe] on the neutron-excess causes [Na/Fe] to be below 0, since for the Virgo dwarfs [Fe/H] is lower than solar ($\sim$ -0.5). The extended star formation history then could also cause considerable Ca-enrichment, leading for the dwarf ellipticals to larger [Ca/Fe] values, since they accrete material from the 
metal-rich cluster environment, rather than the Fornax dwarf, which accretes more pristine gas in the Local Group, which causes lower [Ca/Fe] ratios. An important clue might be strong correlation between [Na/Fe] and [Fe/H], 
when one considers Local Group dwarf galaxies like Fornax and the LMC, dwarf elliptical galaxies, the disk of the Milky Way, and the centers of giant elliptical galaxies.  All this could be due to Na-yields that 
depend strongly on metallicity. The abundance of sodium influences the electron pressure so that the strength of many other features are affected. For example, \citet*{conroyandvandokkum.2012ApJ...747...69C} 
show that for massive galaxies 
an increase in the sodium abundance causes a decrease in the abundance of CaII, which causes a decrease in the EW of CaT. Increasing the sodium abundance can mimic the effects of a more bottom-heavy IMF.
For dwarfs, however, we see a different behavior comparing Ca and Na, although there is no evidence that the IMF is responsible here. And also, for the LMC both [Na/Fe] and [Ca/Fe] have the same sign, implying
that another parameter has to responsible for the difference between LMC and the dEs, presumably the star formation history. However, with this considerable ($\sim$ 0.2) difference in [Ca/Fe] between the LMC and the dEs, it 
is not so obvious that objects like the LMC should be the progenitors of dEs, unless Ca-enrichment by SN Ia in the cluster environment was particularly effective.  
   
What is clear is that the abundance ratio pattern in dwarf ellipticals is very different from that in massive Galactic globular clusters, which show enhanced [Na/Fe], a strong Na-O anticorrelation etc. 
in many stars. It is of course still possible that a fraction of the stars displays these effects, but that fraction is so small that it cannot be detected in integrated light. This difference probably indicates 
that star formation timescales in dwarf ellipticals are long, on the order of Gyrs, since these globular clusters must have been formed on very small timescales, with their ages being so large (\citealt{grattonetal.2001A&A...369...87G}).

\section{Conclusions} 
\begin{itemize}
\item In this paper, we determine abundance ratios of a sample 37 dEs in the Virgo cluster as a part of the SMAKCED project. This sample is representative of the analysis of 
the kinematic properties (\citealp{tolobaetal.2014ApJS..215...17T}) and also all morphological sub-classes found by \citet{liskeretal.2006AJ....132..497L,liskeretal.2006AJ....132.2432L,liskeretal.2007ApJ...660.1186L}. 
\item We use optical spectroscopy to measure a total of 23 Lick indices in the LIS-5 \AA\ flux calibrated system and apply the MILES models to interpret them. We derive new age and metallicity estimates for these galaxies. 
Taking advantage of high resolution spectral data we are able to calculate the abundance ratios of Na and Mg using the models of MILES.
\item We find the unusual behaviour that [Na/Fe] is under-abundant w.r.t. solar. This is opposite to what is found in massive giant elliptical galaxies.
We also find that dEs fall on a relatively tight relation between [Na/Fe] and [Fe/H], which we recently presented in \citet{mentzetal.2016MNRAS.463.2819M}, including Local Group dwarf galaxies, 
the Milky Way and giant elliptical galaxies. From our results, we try to sketch a possible scenario for the evolution of dEs in the Virgo cluster. We find that dEs show disk-like star formation histories favouring them to originate from star forming spirals or dwarfs.
\item  Na-yields appear to be very metal-dependent, in agreement with studies of giant ellipticals, probably due to the large dependence on the neutron-excess in stars.
\item We conclude that dEs have undergone a considerable amount of chemical evolution, they are therefore not uniformly old, but have extended star formation histories, similar to many of the Local Group galaxies.
\end{itemize}
\section*{Acknowledgements}

We thank an anonymous referee for his critical comments, helping to improve the paper. We also thank to F. La Barbera for kindly providing the new models. The research was supported by The Scientific and Technological Research Council of Turkey (TUBITAK) under project number 1059B141401204 and 1649B031406124. 
RFP, AV and JF-B acknowledge support from grant AYA2016-77237-C3-1-P from the Spanish Ministry of Economy and Competitiveness (MINECO). 
Paudel acknowledges the support by Samsung Science \& Technology Foundation under Project Number SSTF-BA1501-0.




\bibliographystyle{mnras}
\bibliography{mbib2} 




\appendix

\section{Ca abundances}

Given the fact that the conversion between Ca4227 and [Ca/Fe] is more uncertain than the ones beween NaD and [Na/Fe] and between Mgb and [Mg/Fe], we have put the discussion of the Ca abundances in the appendix.

For our dEs sample, the measured Ca4227 values are slightly higher than the model predictions for solar-abundance models (Fig.~\ref{fig:index_index_1aug_hgf_ca4227}), implying that Ca is slightly over-abundant in dEs. One might get a very rough idea about the values of [Ca/Fe] when using the only available models, those by CvD for old ages and solar metallicities, to convert Ca4227 to [Ca/Fe]. Although we know that these models are inaccurate, since they must depend on age and metallicity (see the MILES ones), they can serve here to give a crude idea. 
Using this calibration, Figure~\ref{fig:abundances_error_Ca} shows that [Ca/Fe] is a bit larger 
than in the Milky Way disk, but lower than in the thick disk and halo. An interpretation of these high values could that SN Ia's are partially responsible for the Ca-enrichment 
(see \citealp{travaglioetal.2004A&A...425.1029T} and above).

Previous abundance measurements of Ca (\citealt{michielsenetal.2003ApJ...597L..21M}) in dEs show two groups: the ones with CaT* $\sim$ 7\AA\ , higher than expected for solar abundance ratio models, 
and a sample of 3 objects with CaT* $\sim$ 5\AA\ . Such lower values could be obtained by having a range in ages, while including some very young objects. The higher values of 7\AA\ , however, 
can only be explained by [Ca/Fe] values that are somewhat larger than solar, consistent with this paper. Our results are also consistent with the line strengths in the Fornax cluster dwarf 
	NGC 1396 (\citealp{mentzetal.2016MNRAS.463.2819M}). Such high [Ca/Fe] values are not seen in giant ellipticals, where the CaT*  index is lower than expected from stellar population models with 
solar abundance ratios (\citealp{sagliaetal.2002ApJ...579L..13S,cenarroetal.2003MNRAS.339L..12C}). While the latter can be explained by a bottom-heavy IMF, a much easier and more natural explanation 
would be in [Ca/Fe] < 0 for such galaxies (\citealp{cenarroetal.2003MNRAS.339L..12C}, see also \citealp*{conroyandvandokkum.2012ApJ...747...69C}). For dEs the CaT* values are so high that IMF-slopes 
steeper than Salpeter are excluded (\citealp{mentzetal.2016MNRAS.463.2819M}), so solar or super-solar [Ca/Fe] values are unavoidable. Note, however, the difference between the dEs and the LMC, which has 
much lower [Ca/Fe] values, but a similar metallicity. We argue that the difference is due to the enrichment by SN type Ia and different star formation histories.

\citet*{prochaskaetal.2005AJ....130.2666P} suggested that a CN band might be affecting the Ca4227 index in giant ellipticals, but since dEs do not show strong CN bands, the Ca abundance can be 
measured reliably from the Ca4227 index (\citealp{vazdekisetal.1997ApJS..111..203V}).

\begin{figure*}
	\includegraphics[width=0.93\textwidth]{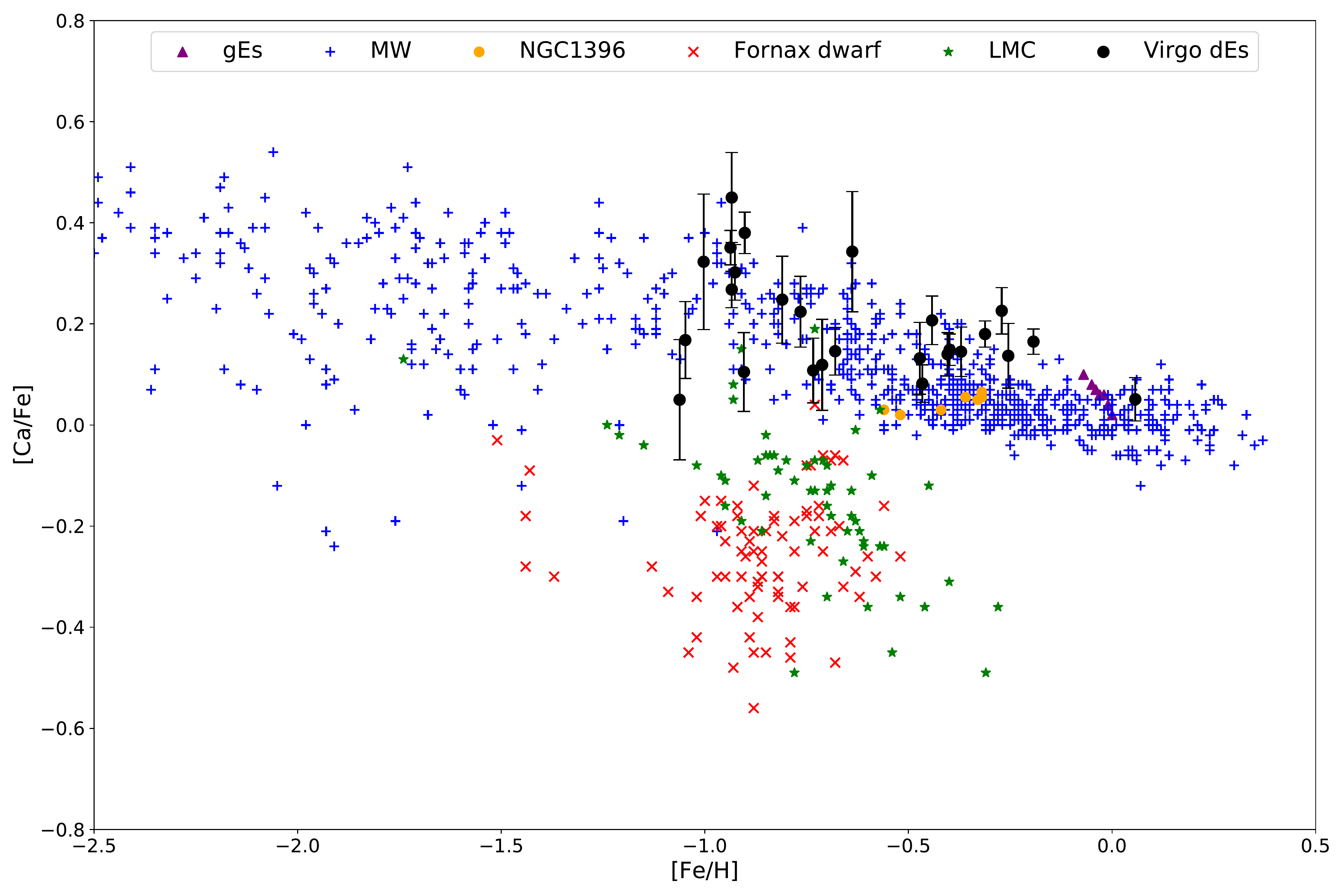}
	\caption{Ca abundances as a function of metallicity [Fe/H]. Description of the points: as in Figure 4.}
	\label{fig:abundances_error_Ca}    
\end{figure*}

\bsp	
\label{lastpage}
\end{document}